\documentclass[12pt]{iopart}
\usepackage{graphicx}

\begin{document}

\title{Thermal transport across metal-insulator interface via electron-phonon interaction}

\author{Lifa~Zhang,$^1$ Jing-Tao L\"u,$^2$ Jian-Sheng Wang,$^1$  Baowen~Li$^{1,3,4}$ }

\address{$^1$ Department of Physics and Centre for Computational
Science and Engineering, National University of Singapore, Singapore
117542, Republic of Singapore}
\address{$^2$ DTU-Nanotech, Department of Micro- and Nanotechnology, Technical University of Denmark (DTU), {\O}rsteds Plads, Bldg.~345E, DK-2800 Lyngby, Denmark}
\address{$^3$ NUS Graduate School for Integrative
Sciences and Engineering, Singapore 117456, Republic of Singapore }
\address{$^4$ NUS-Tongji Center for Phononics and Thermal Energy Science and
Department of Physics, Tongji University, 200092 Shanghai, PR
China
}

\begin{abstract}
The thermal transport across metal-insulator interface can be characterized by electron-phonon interaction through which an electron lead is coupled to a phonon lead if phonon-phonon coupling at the interface is very weak. We investigate the thermal conductance and rectification flowing between the electron part and the phonon part using nonequilibrium Green's function method. It is found that the thermal conductance has a nonmonotonic behavior as a function of average temperature or the coupling strength between the phonon leads in the metal part and the insulator one. The metal-insulator interface shows evident thermal rectification effect, which can reverse with changing of average temperature or the electron-phonon coupling.
\end{abstract}

\pacs{68.35.-p, 
66.70.-f, 
63.20.kd, 
72.10.Di 
}
\maketitle

\section{Introduction}

With the increase of integration density, accumulation of heat becomes a bottleneck for further development of microelectronic devices. Moreover, as most of electronic devices consist of metal and insulator/semiconductor interfaces, to understand the thermal transport through the metal-insulator/semiconductor interface is indispensable for heat dissipation \cite{vanka01,cahill03,costescu04}.

\begin{figure}[t]
\includegraphics[width=0.8\columnwidth]{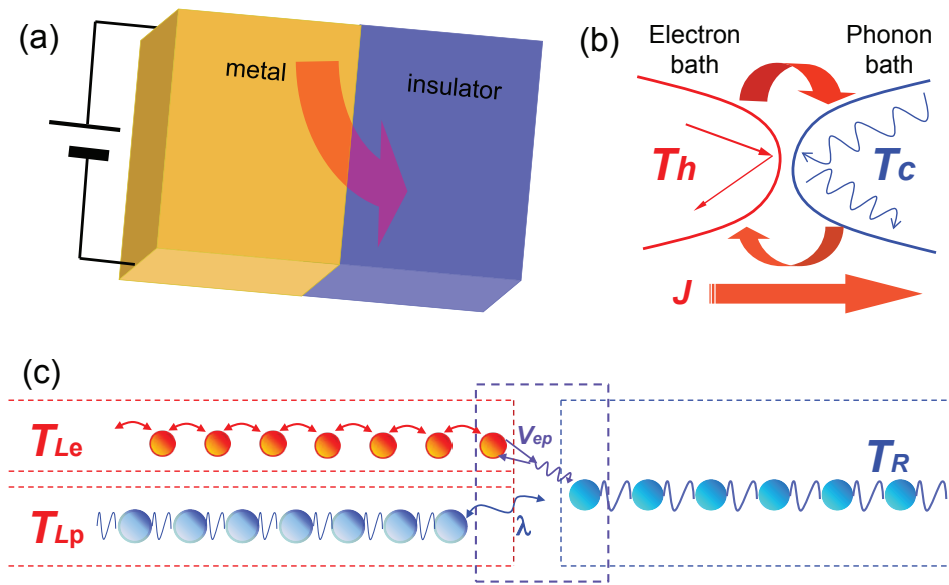}
\caption{\label{fig1_model}(a) Heat generation in a metal with an applied bias. The arrow shows the heat transport from the metal to the insulating substrate. (b) An electron bath and a phonon bath at two different temperatures connected by EPI. Energy exchanges between the two baths and there is a net thermal current $J$ from the hot bath to the cold one. (c) The lattice model of the metal-insulator interface. The one dimensional electron lead (the left upper semi-infinite chain where the dots represent electron states ) connects to the one dimensional atomic chain (the right semi-infinite chain) via electron-phonon interaction $V_{\rm ep}$. The phonon part in the metal (the left lower semi-infinite atomic chain) can also be considered to the thermal transport, which connects to the right atomic chain by a relative coupling $\lambda$. }
\end{figure}

In metals electrons dominate thermal transport while phonons do in
semiconductors and insulators; thus for thermal transport across metal-nonmetal
interfaces, energy must transfer between electrons and phonons
\cite{majumdar04}. Thanks to many remarkable physical properties and prospects
for large-area epitaxial growth, graphene is a promising material for future
electronics \cite{castro11}. The biased graphene on a dielectric insulator
often appears in electronic devices especially in transistors. Due to the bias the electrons in metallic part have a higher effective temperature than the phonons in the insulating substrate; and the thermal transport from the electrons in metal to the phonons in insulator is one important channel of energy transfer \cite{chen08,pere09}. It is found in a very recent experiment on a
carbon nanotube that more that 80 percent of the electrical power supplied to
the nanotube is transferred directly into the substrate by the electron-phonon interaction (EPI)
\cite{balo12}. Therefore to understand the thermal transport across the
interface through the EPI is highly desirable for the heat dissipation in
electronic devices.

To study the thermal transport across interface, the acoustic mismatch model \cite{little59} and the diffuse mismatch model \cite{swartz89, majumdar04} are widely applied; however, both models offer limited accuracy in nanoscale interfacial resistance predictions \cite{stevens05} because they neglect atomic details of actual interfaces. A scattering boundary method within the lattice dynamic approach \cite{lumpkin78,wang06,zhang08} fully considers the atomic structures in the interface; but it can only be applied to ballistic thermal transport.  Classical molecular dynamics is another widely used method in phonon transport \cite{li05,term09,landry09}, which is not accurate below the Debye temperature, and ignores the quantum effect.  Only recently the nonequilibrium Green's function method, which originates from the study of electronic transport \cite{haug96},  been applied to study the quantum phonon transport \cite{zhang07,wang08,zhang11, zhang09b}. So far the study of the coupled electronic and phonon transport \cite{lu07,galperin07,musho11} is rare, especially in the metal-nonmetal interface \cite{majumdar04,pop07,ordonez11}.

In this paper, using the nonequilibrium Green's function method, we study the thermal transport across metal-insulator interface via the EPI. Our model can also be applied to metal-semiconductor interface. We study energy flow between an electron bath and a phonon bath which are connected by electron-phonon coupling, as shown in Fig.\ref{fig1_model} (b). The electron bath, described by a semi-infinite electronic chain under tight-binding approximation, connects with a phonon lead illustrated as a semi-infinite harmonic atomic chain by a weak electron-phonon interaction, as shown in Fig.\ref{fig1_model} (c), where the phonon degrees of freedom in metal can also be considered in the thermal transport.

\section{Model and Method}
We study the interfacial thermal transport at the metal-insulator interface as shown in Fig.~\ref{fig1_model}(a). To manifest the effect of the interface we exclude the nonlinear electron or phonon transport in the two materials themselves, thus the only thermal resistance comes from the interface. Such model really uncovers the thermal transport properties of the interface itself. To study the longitudinal thermal transport, that is, the cross-plane interfacial transport, we can simplify the problem further to thermal transport between an electron bath and a phonon bath which are connected by an electron-phonon coupling, as shown in Fig.~\ref{fig1_model}(b), and can be represented by one dimensional lattice model as shown in Fig.~\ref{fig1_model}(c), where a semi-infinite electronic chain (at temperature $T_{Le}$) connects to a semi-infinite atomic chain (at temperature $T_{R}$) by an electron-phonon interaction $V_{ep}$. The phonon part in the metal can also be considered by another semi-infinite atomic chain which connects to the right atomic chain by a relative coupling $\lambda$.  In our model, the scattering for both electrons and phonons only comes from the interface, and the electron and phonon transport in the corresponding semi-infinite periodic leads is ballistic. Therefore we can partition the system into three parts ($L$, $C$, $R$), the atoms at the interface are regarded as center illustrated as the dashed-line rectangle in Fig.\ref{fig1_model} (c), and the rest parts are leads. For simplicity, we only consider one dimensional case, ignoring electron spin. Thus the Hamiltonian of the whole system in Fig.\ref{fig1_model} (c) is written as
\begin{equation}
H=H^L_{\rm e}+H^C+H^{LC}_{\rm e}+ \sum_{\alpha=L,R}(H^{\alpha}_{\rm p}+ H^{\alpha C}_{\rm p}),
\label{eq:ham0}
\end{equation}
where $H^L_{\rm e}=\sum_i\varepsilon_0c^\dagger_i c_i-\sum_{|i-j|=1} t c^{\dagger}_ic_j$ is the Hamiltonian of the electron lead in the left part. The electron coupling between the left lead (the site 1) and the center is $H^{LC}_{\rm e}=-t c^{\dagger}_1 c-t c^{\dagger} c_1$. We set atomic mass $m=1$, and the Hamiltonian of the center is
\begin{equation}
	H^C = \frac{1}{2} \dot{u}^2 + \frac{1}{2}(\lambda k^L +k^R)u^2 +\varepsilon_0c^\dagger c+ V_{\rm ep}c^\dagger c u,
	\label{eq:hamc}
\end{equation}
where $k^L$ and $k^R$ are the spring constants of the left and right atomic chains, respectively. $u$ is the atom displacement. $\lambda k^L$ denotes the coupling between the two atomic chains; $\lambda$ can be chosen from 0 to 1. $V_{\rm ep}$ is the electron-phonon coupling. The phonon Hamiltonian of two leads and its coupling to the center are
$ H^\alpha_{\rm p} = \frac{1}{2}\sum_{i} \dot{u}^\alpha_i\dot{u}^\alpha_i + \frac{1}{2}\sum_{|i-j|=0,1} u^\alpha_i K^\alpha_{ij} u^\alpha_j$ and $
	H^{\alpha C}_{\rm p} = u^{\alpha}_1 K^{\alpha C}u$ \cite{wang08,lu07}.

Applying the standard procedure of nonequilibrium Green's function method \cite{wang08},  without the EPI, we obtain the electron retarded Green's function  $G^r_0(\varepsilon) = \left[\varepsilon - \varepsilon_0- \Sigma^{r}_L(\varepsilon)\right]^{-1}$, where  $\Sigma^{r}_L= g^{r}_L t^2$ is the retarded self-energy with a surface Green's function $g^r_L$, where $g^{r}=\left[(\varepsilon + i\eta)I - H^L_{\rm e}  \right]^{-1}$. The less Green's function can also be easily obtained. The
phonon retarded Green's functions is $D^r_0(\omega) = {D^a_0}^\dagger(\omega) = \left[\omega^2-(\lambda k^L +k^R)-\Pi^r_L(\omega)-\Pi^r_R(\omega)\right]^{-1}$, where
$\Pi^r_L(\omega)=(\lambda k^L)^2 d^{r}_L(\omega)$ and $\Pi^r_R(\omega)=(k^R)^2 d^{r}_R(\omega)$ are the retarded self energies of left and right leads with surface Green's function $d^{r}_L$ and $d^{r}_R$, where $d^{r}_{L,R}=\left[(\omega + i\eta)^2 I - K^{L,R} \right]^{-1}$. The EPI is included as perturbation. The full Green's functions are obtained from the Dyson equation, that is, $G^{r,a} =
[(G_0^{r,a})^{-1}-\Sigma^{r,a}_{\rm ep}]^{-1}$, and $G^{<,>} = G^r (\Sigma^{<,>}_{\rm ep}+\Sigma^{<,>}_{L})
G^a$, where $\Sigma^{r,a,<,>}_{\rm ep}$ is the self-energy from the EPI. For phonon, we have $D^{r,a} =
[(D_0^{r,a})^{-1}-\Pi^{r,a}_{\rm ep}]^{-1}$, and $D^{<,>} = D^r (\Pi^{<,>}_{\rm ep}+\Pi^{<,>}_{L}+\Pi^{<,>}_{R}) D^a$, where the nonlinear self energy $\Pi^{r,a,<,>}_{\rm ep}$ comes from the electron-phonon coupling.  Keeping the lowest non-zero order (second order) of the self-energies, we could obtain the nonlinear self-energies $\Sigma^{r,a,<,>}_{\rm ep}$ and $\Pi^{r,a,<,>}_{\rm ep}$  up the second order, which are general for any dimensional systems. For our one-dimensional  simple model as shown in Eq.\ref{eq:hamc}, the self-energies could be written as
\begin{equation}
	\Sigma_{\rm ep}^{>,<}(\varepsilon) =i V_{\rm ep}^2  \int {G_0^{>,<}}(\varepsilon-\omega){D_0^{>,<}}(\omega)\frac{d\omega}{2\pi},
	\label{eq_epsl}
\end{equation}
and
\begin{eqnarray}
	\Sigma_{\rm ep}^{r}(\varepsilon)&=& iV_{\rm ep}^2 \bigl\{ -{D_0^r}(\omega'=0)\int {G_0^<}(\varepsilon')\frac{d\varepsilon'}{2\pi} \nonumber \\
 && + \int \frac{d\omega}{2\pi} \bigr[{G_0^r}(\varepsilon-\omega){D_0^<}(\omega) +{G_0^<}(\varepsilon-\omega){D_0^r}(\omega) \nonumber \\
 &&  + {G_0^r}(\varepsilon-\omega){D_0^r}(\omega)\bigr] \bigr\}.
	\label{eq_epsr}
\end{eqnarray}
The nonlinear self-energies for the phonons are
\begin{equation}
	\Pi_{\rm ep}^{>,<}(\omega) = -iV_{\rm ep}^2 \int \frac{d\varepsilon}{2\pi}{G_0^{>,<}}(\varepsilon){G_0^{<,>}}(\varepsilon-\omega),
	\label{eq_eppl}
\end{equation}
and
\begin{eqnarray}
	\Pi_{\rm ep}^r(\omega)&=&-i V_{\rm ep}^2 \int \frac{d\varepsilon}{2\pi}\bigl[{G_0^r}(\varepsilon){G_0^<}(\varepsilon-\omega) \nonumber \\
 && +{G_0^<}(\varepsilon){G_0^a}(\varepsilon-\omega)\bigr].
	\label{eq_eppr}
\end{eqnarray}
Eqs.~(\ref{eq_epsl}-\ref{eq_eppr}) are the so-called Born approximation (BA). By replacing the bare Green's functions $G_0$ and $D_0$ with the full Green's functions $G$ and $D$, we do iteration under self-consistent Born approximation (SCBA). While the BA fails to satisfy the energy current conservation, the SCBA fulfills it \cite{lu07}.

Therefore, we obtain the heat current from electron part as \cite{haug96,meir92,jauho94}
\begin{equation}
	J^{\rm e} =  \int \frac{d\varepsilon}{2\pi\hbar} \varepsilon~{\rm Tr}\{G^>(\varepsilon) \Sigma^<_L(\varepsilon) - G^<(\varepsilon)\Sigma^>_L(\varepsilon)\}.
	\label{eq_je}
\end{equation}
The heat current from the phonon lead is \cite{wang08,lu07,wang07}
\begin{equation}
	J^{\rm p}_\alpha = -\int \frac{d\omega}{4\pi} \hbar\omega~ {\rm Tr}\{D^>(\omega) \Pi^<_\alpha(\omega) - D^<(\omega)\Pi^>_\alpha(\omega)\}.
	\label{eq_jp}
\end{equation}
Due to energy conservation, $J^{\rm e}+J^{\rm p}_L+J^{\rm p}_R=0$.
If the coupling $\lambda=0$, we obtain $J^{\rm e}=-J^{\rm p}_R$ when the energy flows across the interface only through EPI, that is, the heat generation of the metal part transfers directly to the insulator part. We set the Plank constant $\hbar=1$ and the Boltzman constant $k_{\rm B}=1$ in the following numerical calculation; and $\varepsilon_0=0$, $k^L=k^R=1,\;t=1$. We define the conductance $\sigma=|J^{\rm e}|/\Delta T$ to illustrate the thermal transport between the electron part and phonon part.

\section{Thermal Conductance across the Interface}
\begin{figure}[t]
\includegraphics[width=0.8\columnwidth]{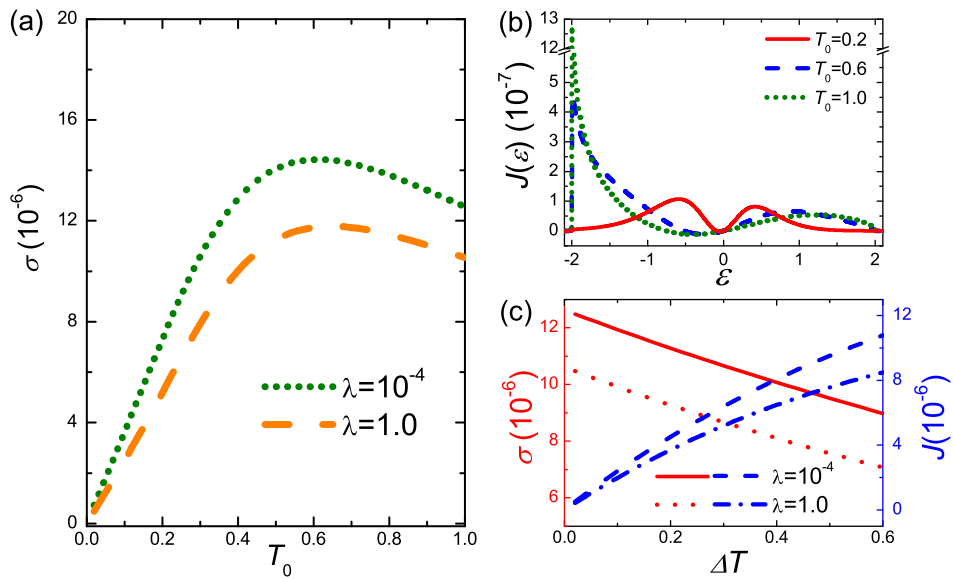}
\caption{\label{fig2_cond} (a) Thermal conductance across the metal-insulator interface vs the average temperature of the baths. $T_{L \rm e}=T_0+\Delta T$, $T_{L \rm p}=T_{R}=T_0-\Delta T$, $\Delta T=0.01$, $V_{\rm ep}=0.01$. (b) The thermal current density $J(\varepsilon)$ ($J^{\rm e}=\int d\varepsilon J(\varepsilon)$) for electrons with energy $\varepsilon$ at $\lambda=0^+=10^{-4}$.  (c) The thermal conductance (left scale) and current (right scale)  from the electron bath  vs the temperature difference $\Delta T$ at $T_0=1.0$. }
\end{figure}

As shown in Fig.\ref{fig1_model}(a), if the metal part is applied  by a voltage,
the average energy of electrons increases, which means the electrons have a
higher effective temperature than the phonons. Thus we set the left and right
phonon leads to a lower temperature $T_{L \rm p}=T_{R}=T_0-\Delta T$,
while the electron lead has a higher temperature $T_{L \rm e}=T_0+\Delta T$.
If we don't consider the phonon contribution in the metal part, the thermal
conductance from the electron lead to the phonon lead has a nonmonotonic
behavior with increasing the average temperature  $T_0$, as shown in
Fig.~\ref{fig2_cond}(a). In the low-temperature region, as temperature
increases, the thermal conductance increases due to more contribution from
electrons with energy far from Fermi surface ($\varepsilon=0$) although there is certain
decrease from the contribution near Fermi surface, which can be seen from the
solid line ($T_0=0.2$) and dashed line ($T=0.6$) in Fig.~\ref{fig2_cond} (b).
With the temperature further increasing, the dominating effect comes from
the decrease of the contribution near the Fermi level, as shown in the dashed
line and dotted line of Fig.~\ref{fig2_cond} (b). This causes the thermal
conductance to decrease after certain temperature.  Such dependence of thermal
conductance on temperature is consistent with the experiment observation of
thermal contact conductance between graphene and silicon dioxide \cite{chen09}.
 While the conductance has a nonmonotonic dependence
on $T_0$, the heat current and conductance shows a monotonic behavior with
temperature difference $\Delta T$ increasing, as shown in the inset of
Fig.~\ref{fig2_cond} (c).
If we include the phonon contribution in the metal part, the thermal
conductance has a similar temperature dependence, but with
a smaller magnitude. This means that the heat generation from the electron
part is affected by the coupling strength between the two phonon parts in the
metal-insulator interface.

\begin{figure}[t]
\includegraphics[width=0.8\columnwidth]{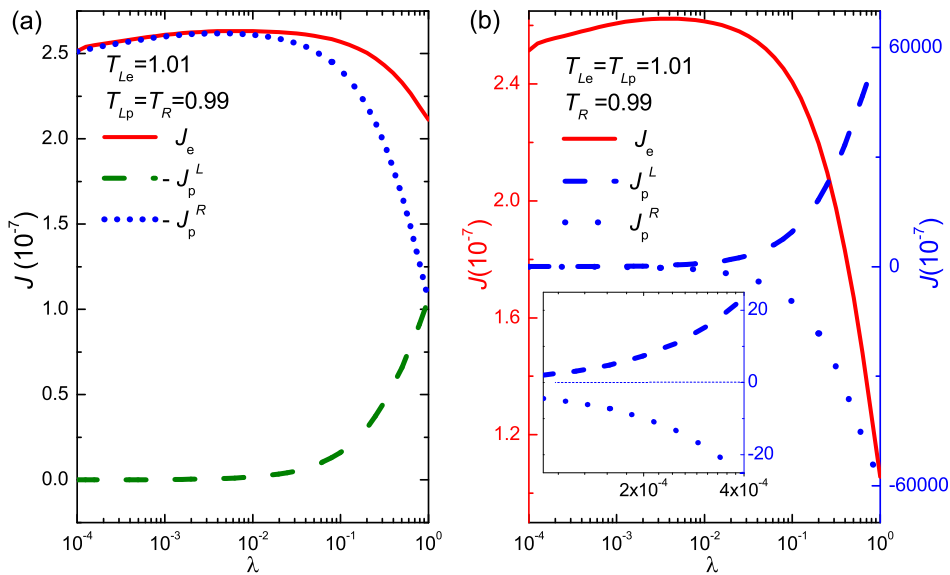}
\caption{\label{fig3_heat} (a) Thermal current vs the coupling $\lambda$. The solid line corresponds to the thermal current $J^{\rm e}$ flowing from the electron bath. The dashed and dotted lines correspond to the thermal current ($-J^{\rm p}_L$ and $-J^{\rm p}_R$ ) flowing into the left and right phonon baths, respectively. (b) Thermal current vs the coupling $\lambda$ at $T_{L \rm e}=T_{L \rm p}=1.01$, $T_{R}=0.99$.  The solid, dashed and dotted lines correspond to the thermal current flowing from the electron bath (left scale), the left and the right phonon baths (right scale), respectively. The inset is the zoom-in of (b) for small $\lambda$.  For (a) and (b), $V_{\rm ep}=0.01$. }
\end{figure}

In order to investigate the effect of phonon-phonon channel $\lambda$ on the thermal transport through electron-phonon channel $V_{ep}$, in
Fig.~\ref{fig3_heat} (a) we plot the heat current from the electrons to the phonons when we increase the coupling strength $\lambda$ between the two phonon parts.
We find that in the small $\lambda$ region, with $\lambda$ increasing the heat
current flows into the right phonon lead will increase due to the smaller
difference between the atom ($K^C=(\lambda k^L +k^R)$) in interface and the
atoms ($K^R_{ii}=2k^R$) in the right lead, thus the scattering for phonons at
the interface decreases. The heat is easier to transport to the right
lead; $J^{\rm e}$ increases. At the same time, the heat can flow from the electrons to the left phonon lead; $-J^{\rm p}_L$ increases with
$\lambda$. When the $\lambda$ increases further, the phonon scattering between left chain and the center atom decreases, more thermal energy transferred from electron is easier to flow into the left lead. Such fast increasing of $-J^{\rm p}_L$ causes the decreasing of the heat into the right lead $-J^{\rm p}_R$. With increasing of $\lambda$, the phonon scattering at the interface between the left and right atomic chains decreases, more phonons can transport coherently in the two atomic leads, and are less efficient to couple to electrons; thus the heat flowing from the electron lead begins to decrease. Due to continuous increase of the heat flowing into the left phonon lead $-J^{\rm p}_L$, and $J^{\rm e}+J^{\rm p}_L+J^{\rm p}_R=0$, thus the decrease of heat from electrons $J^{\rm e}$ lags behind the decrease of the heat into the right lead $-J^{\rm p}_R$. And if $\lambda=1$,
the heat into two phonon leads has the same value due to the symmetry.   As
shown in Fig.~\ref{fig3_heat} (b), if the non-biased metal part
in a higher temperature while the insulator in a lower one, the heat flow from
electron part has a similar dependence on $\lambda$; but the heat flowing
from left phonon lead and the one into the right phonon lead monotonically
increase with  $\lambda$ since phonon transport dominates heat flow in such
case and the phonon scattering decreases. We also find that, in this case the thermal transport through the phonon-phonon channel $\lambda$ is several orders of magnitude larger than that through the electron-phonon channel. And even if $\lambda=4\times10^{-4}$ (see the inset of Fig.~\ref{fig3_heat}(b)), the thermal current of phonon is ten times larger than that of electron. Therefore for the thermal transport in the non-biased metal-insulator interface, such small thermal current of electron can be ignored, which is consistent with the recent finding in Ref.\cite{sing13}. However, for the interface between a biased metal and a insulator, the thermal transport through the electron-phonon interaction is very important, which dominated the thermal transport even if $\lambda=1$ as shown in Fig~\ref{fig3_heat} (a), which can explain the remote Joule heating via electron-phonon interaction in Ref. \cite{balo12}.

\section{Thermal Rectification across the Interface via EPI}
\begin{figure}[t]
\includegraphics[width=0.8\columnwidth]{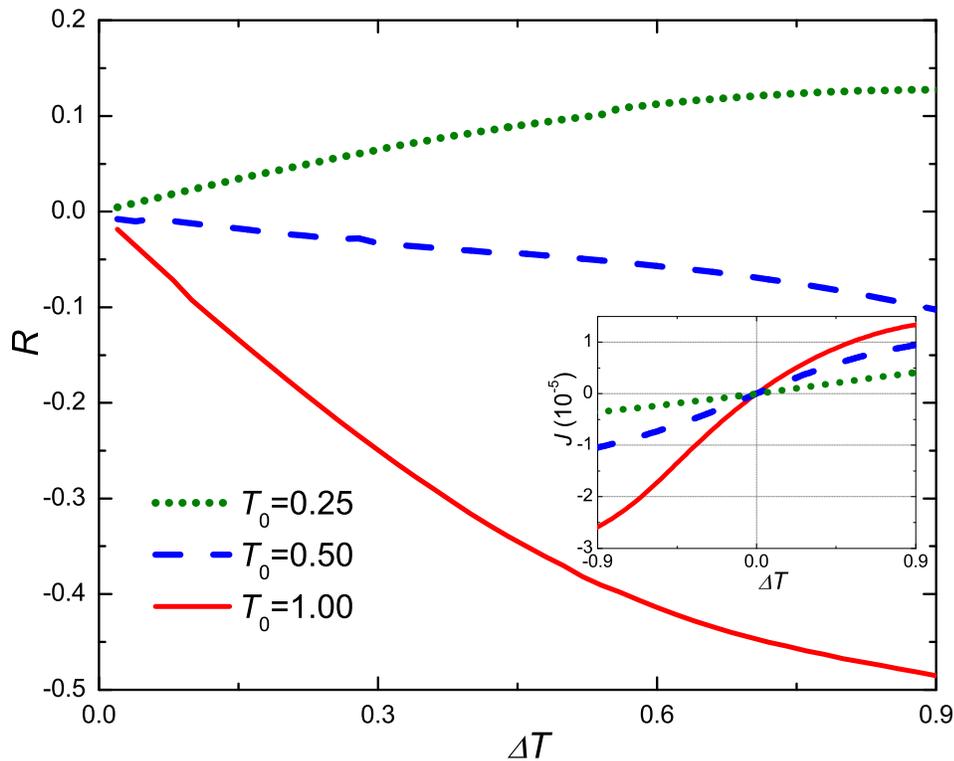}
\caption{\label{fig4_rect1} Thermal rectification of the metal-insulator interface vs temperature difference. The inset: the thermal current across the interface vs temperature difference. $T_{h}=T_0(1+\Delta T)$, $T_{c}=T_0(1-\Delta T)$, $V_{\rm ep}=0.01$. }
\end{figure}
Thanks to nonlinearity and asymmetry of the metal-insulator interface, we can
expect the thermal rectification, which is defined as $R =(J_+ - J_-)/ {\rm max
}\{J_+, J_-\}$, where $J_+$ is the forward direction heat flux, defined as $T_L=T_h, T_R=T_c$,
and $J_-$ is that of the backward direction when $T_L=T_c, T_R=T_h$. Here, $T_h$ and $T_c$
correspond to the temperatures of the hot and cold baths, respectively.
In the rest of the paper, we will not consider the phonon contribution in metal (we set $\lambda=10^{-4}$, $T_{Lp}=T_R$ in the calculation to avoid divergence), which
will not change the physical properties of the rectification, and only decrease
its magnitude. Figure~\ref{fig4_rect1} shows the dependence of thermal
rectification on temperature difference for different average temperatures. The
rectification shows a monotonically increasing behavior of temperature
difference due to the bigger difference of heat current in larger temperature
difference as shown in the inset of Fig.~\ref{fig4_rect1}, which is consistent
with all the traditional study on thermal rectification \cite{li04,li12}. With
the temperature $T_0$ increasing, we find that the rectification can change
sign.

\begin{figure}[t]
\includegraphics[width=0.8\columnwidth]{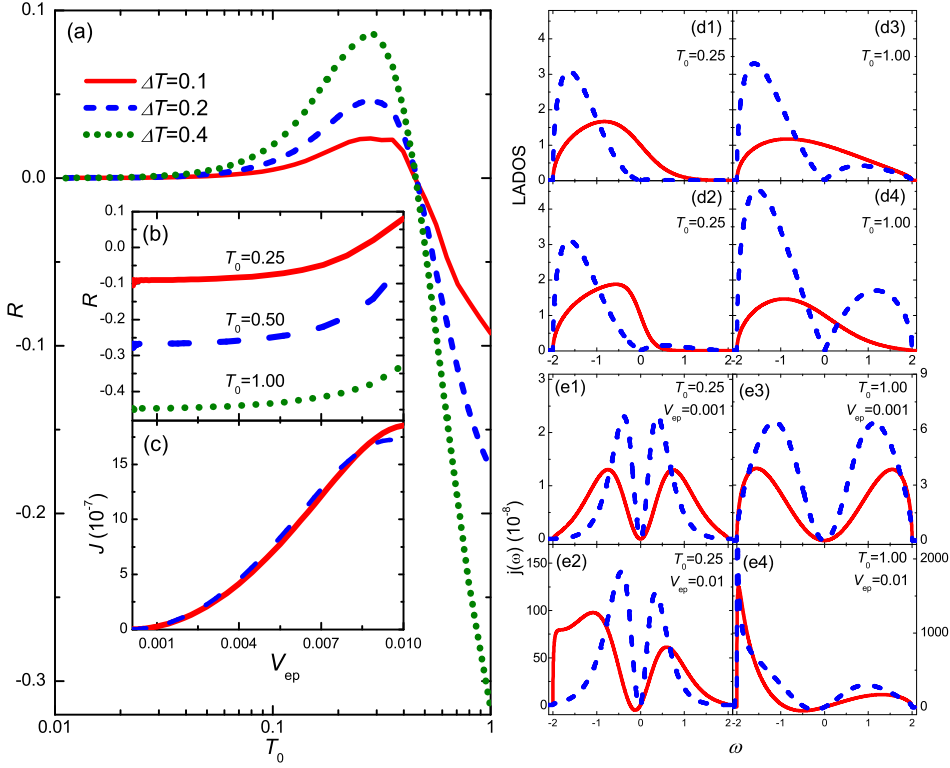}
\caption{\label{fig5_rect2} (a) Thermal rectification of the metal-insulator interface vs temperature $T_0$ for different temperature differences at $V_{\rm ep}=0.01$. (b) Thermal rectification vs the electron-phonon interaction $V_{ep}$ at different temperatures. (c) The forward thermal current (solid line) and backward thermal one (dashed line) vs $V_{\rm ep}$ at $T_0=0.25$. (d1)-(d4) The local available density of states (LADOS)  at surface for the left electron part ($\rho_{\rm e}={|\rm Img}(g^<)|$)(solid line) and the right phonon part ($\rho_{\rm p}={|\rm Img}(d^<)\omega|$)(dashed line). (d1)((d2)) and (d3)((d4)) are corresponding to the forward (backward) transport at $T_0=0.25$ and $1.0$, respectively. (e1)-(e4) The forward (solid lines) and backward (dashed lines) thermal current density $J(\varepsilon)$ for different temperatures and different EPI.   For all the curves, $T_{h}=T_0(1+\Delta T)$, $T_{c}=T_0(1-\Delta T)$; from (b) to (e) $\Delta T=0.4$.}
\end{figure}

The dependence of thermal rectification on temperature and the EPI is shown in
Fig.~\ref{fig5_rect2}. We find that at a relative larger EPI ($V_{\rm ep}=0.01$),  with increasing the temperature, the thermal
rectification can change sign from positive to negative.  However for a very weak EPI, the rectification always keeps negative, as shown in
Fig~\ref{fig5_rect2}(b). Within the range of $\lambda=10^{-4} \sim 10^{-2}$, the rectification is always negative for higher temperatures,  while it can change sign from negative to positive in a lower temperature $T=0.25$.  The relation of thermal currents in the forward and backward direction changes with increasing of $\lambda$ while both their magnitudes monotonically increase, as shown in Fig~\ref{fig5_rect2}(c).

From the local available density of states (LADOS) of electrons and phonons at the surface of the corresponding
leads, we find that at a higher temperature $T_0=1.0$, going from the forward transport to the backward one, the main change is that the LADOS phonons largely increases, as shown in Fig.~\ref{fig5_rect2} (d3) and (d4), which causes a larger current in the backward direction than that in the forward direction, thus the rectification is negative. At a lower temperature, the change of LADOS for electrons and phonons are not obvious, as shown in Fig.~\ref{fig5_rect2} (d1) and (d2), and the rectification could change with the strength of EPI.

Turning on the EPI, the electron Green's functions change much more than the phonon ones; with increasing EPI, the Green's functions of electrons $G^>$ and $G^<$ become more asymmetric, especially for the $G^<$ which have larger value at the energy far away from the Fermi level ($\varepsilon=0$).  At a very weak EPI, the heat current comes from the contribution of electronic energies symmetrically away from the Fermi level. The heat current of the backward transport is larger than that of the forward one, as shown in  Fig.~\ref{fig5_rect2} (e1) and (e3), due to the increase of the LADOS
of phonons in the backward direction, thus the rectification are negative for all the temperatures at weak EPI as shown in Fig~\ref{fig5_rect2}(b). With a stronger EPI, which induces larger value of $G^<$ at the energy range $-2\sim -1$, where the center electron has larger density of states. Thus we find that the forward current has larger value at the range of $-2\sim -1$ as shown in Fig.~\ref{fig5_rect2}(e2). At the same time, the heat current can be affected by the temperature, in the low temperature range, as temperature increases, more electrons with energy far away from Fermi level contribute to the thermal current, as discussed in Fig.~\ref{fig2_cond}(b). Thus in Fig.~\ref{fig5_rect2}(e2), the forward heat current is larger than the backward one; the rectification is positive at low temperatures with larger EPI. In the higher temperature range, the contribution from energy far away from Fermi level will decrease with temperature increasing as shown in Fig.~\ref{fig2_cond}(b); further more, the mainly change from LADOS of electron and phonon is the larger phonon LADOS in the backward transport than that in the forward one; thus the backward heat current is larger than the forward one as shown in Fig.~\ref{fig5_rect2}(e4). Therefore, due to mismatch of electron LADOS and phonon one, the rectification is negative for high temperatures or weak EPI.  Due to the nonlinearity of EPI and its relation with temperature, we can find the positive rectification with relative larger EPI at low temperatures.
\section{Discussions}
\begin{figure}[t]
\includegraphics[width=0.8\columnwidth]{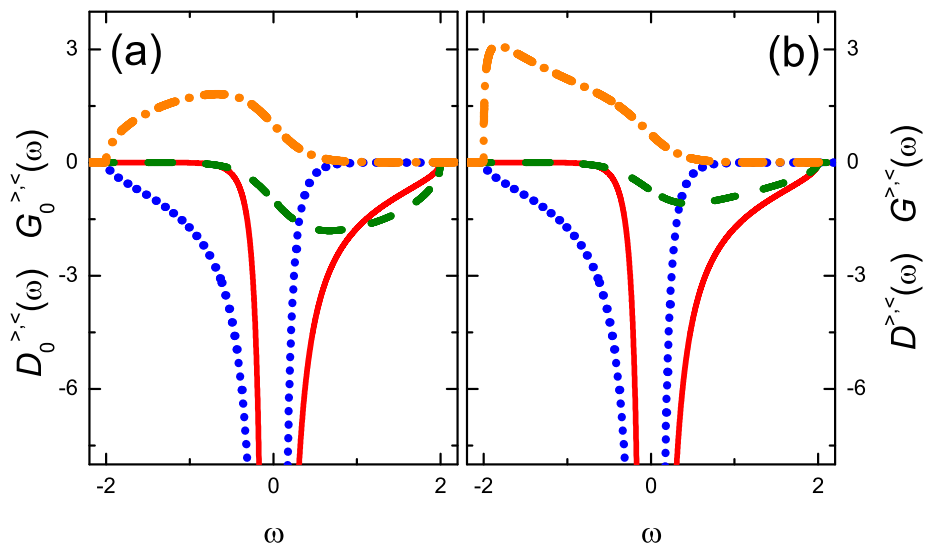}
\caption{\label{fig6_green}  Green's functions for electrons and phonons,  $T_{L \rm e}=T_0+\Delta T$, $T_{L \rm p}=T_{R}=T_0-\Delta T$, $T_0=0.2$, $\Delta T=0.01$. (a) The imagine part of the bare greater and less Green's functions for phonons and electrons: $D_0^>$ (solid line),$D_0^<$ (dotted line),$G_0^>$ (dashed line),$G_0^<$ (dash-dotted line). (b) The imagine part of the full Green's functions for phonons and electrons: $D^>$ (solid line),$D^<$ (dotted line),$G^>$ (dashed line),$G^<$ (dash-dotted line) with the EPI ($V_{\rm ep}=0.01$).  }
\end{figure}
\subsection{Full Green's functions in the calculation}
Under the SCBA, we calculate the self energies of the EPI; after the iteration is convergent we obtain all the Green's functions of  $G^{>,<}$ and $D^{>,<}$ in Eq.~(3) and (4) in the main text, then we can calculate the heat currents. If we do not consider the phonon contribution in the metal part, we should choose the coupling  $\lambda$ be a very tiny positive value, thus the heat current to the left part will be very small and negligible; as shown in Fig.~3 in the main text, if $\lambda=10^{-4}$ the heat current flowing into the left phonon leads is almost zero. If $\lambda=0$, the iteration to calculate self energies for EPI will be divergent, but if $\lambda=0^+=10^{-4}$, it can be convergent. Thus here the tiny $\lambda=0^+$ plays a role as very small onsite potential for the center atom, which is similar as the tiny onsite in computing the phonon Hall effect \cite{wang09}.
We plot the bare Green's functions $G_0$, $D_0$ and the full Green's functions $G$, $D$ in Fig.~\ref{fig6_green}. Without the EPI, the bare Green's functions of electrons $G_0^<$ and $G_0^>$ are symmetric with respect to the origin as shown in Fig.~\ref{fig6_green} (a), but which symmetry for $G^<$ and $G^>$ will be broken and change much if we turn on the EPI, which can be seen in Fig~\ref{fig6_green} (b). However, the Green's functions for phonons do not change much.

\subsection{The role of Fermi energy of electrons}
\begin{figure}[t]
\includegraphics[width=0.8\columnwidth]{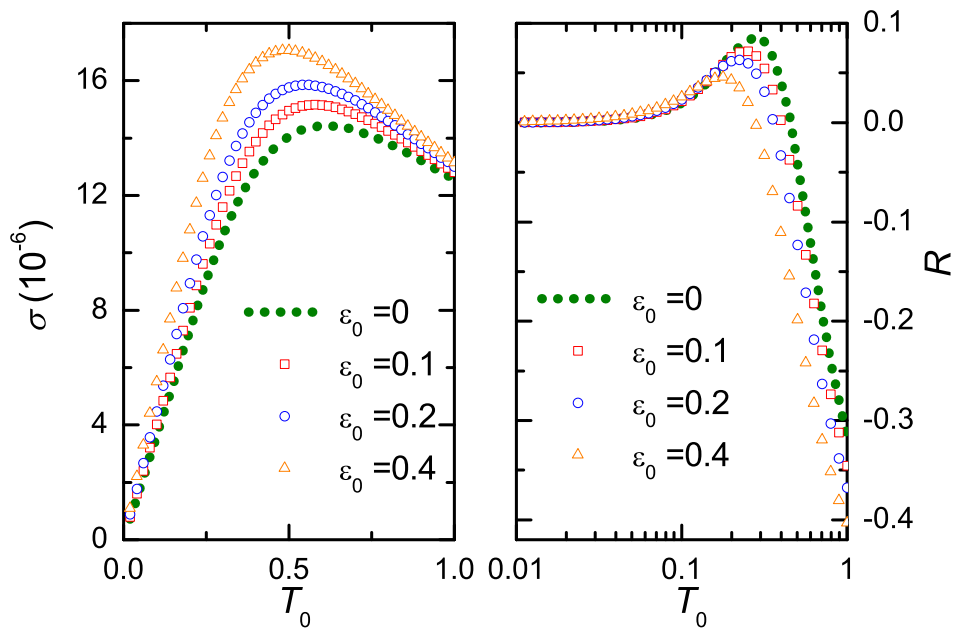}
\caption{\label{fig7_fermi} The conductance (left panel) and the rectification (right panel) vs the onsite $\varepsilon_0$.  $\lambda=10^{-4}$. For comparison, The dotted lines in (a) and (b) are copied from the corresponding line in Fig.2 (a) and Fig.5(a), respectively.    }
\end{figure}

We set the onsite $\varepsilon_0=0$ in the main text, thus the Fermi energy of the electron system is $\varepsilon=0$. However we can change the onsite $\varepsilon$, as shown in Fig.~\ref{fig7_fermi}, the onsite potential can change the magnitude of the thermal conductance and the rectification, and the temperature where the thermal conductance and rectification have maximum value shifts a bit. However the curves of the thermal conductance and rectification have the same properties with the case at $\varepsilon_0=0$. Therefore, the mechanism found for the non-monotonicity of conductance and reversal of rectification still holds for the case of nonzero Fermi energy.
\subsection{Limitation and application}
We use a simplified one-dimensional model to study thermal conductance and rectification across the metal-insulator interface via electron-phonon interaction. For some quasi-one dimensional metal-insulator interface systems, the thermal transport can simply be recast to one-dimensional model, then our results of the conductance and rectification can be applied. The thermal resistance of a whole interface system comes from the two materials themselves and the interface between them. Our study in this work focuses on the interface itself, which provides understanding on the interface in the thermal transport. Although the thermal transport via electron-phonon interaction can be ignored in the non-biased metal-insulator interface, it dominates the thermal transport in the remote Joule heating in a biased metal-insulator interface.

For real materials, there would be more atoms in one unit cell, thus more phonon branches will contribute to the thermal transport. To compare with experiments on real materials, the simplified model need to be generalized to two-dimensional or three dimensional with input of the real parameters.  For realistic three dimensional systems,  the density of states
of electrons and phonons are quite different from those in one dimension \cite{asay08}, then the analysis for the mechanism of the thermal rectification would be different and more complicated; but the general properties on the interfacial thermal transport, such as thermal conductance as functions of $T$, $\lambda$ and the rectification tuned by temperature or electron-phonon interaction, would still hold and could be verified by the future experiments.

The nonequilibrium Green's function approach is a good candidate to solve the interfacial thermal transport.  If we study two or three dimensional systems with large number atoms in the interface, the self-consistent process is time consuming, and we also need to pay more attention to its convergency. If we only are interested on the thermal conductance accurate to the second order of $V_{ep}$, the Born approximation is applicable to avoid the convergent problem. However, for the thermal rectification as a function of $V_{ep}$ can only be obtained by the self consistent Born approximation. Therefore the calculation of the simple model can provide the understanding of the basic properties of thermal transport at metal-insulator interface and can be generalized to real materials with the parameters obtained from first-principle calculation. The challenges would come from the time consuming for the self-consistent calculation of the nonlinear self energies $\Sigma_{\rm ep}$ and $\Pi_{\rm ep}$.
\section{Conclusion}
In summary, using nonequilibrium Green's function method we study the
thermal transport across metal-insulator interface with an EPI. The thermal
conductance and rectification across the interface is thoroughly investigated
under SCBA by a clean and efficient model where an electron lead couples to a
phonon lead through the EPI. We find that the thermal conductance is a
nonmonotonic function of average temperature, it has a maximum value at certain
temperature. Considering the phonon part in metal contributing to the thermal
transport, the heat flow from electron part will change while it has a similar
dependence on temperature.   The thermal rectification effect across the
metal-insulator interface can reverse with varying of system average
temperature and the EPI. Our results are very helpful to explain and
guide the experiment on the thermal transport and dissipation in electronic devices and interface system;  and could be verified in the biased and non-biased graphene sheet supported on insulating substrate. The reversal rectification is significant for the study of one-way heat transport and tuning the direction of thermal transport; thus it could have wide application in the
energy science.

\section*{Acknowledgements}
L. Z.and B. L. are supported by the grant
R-144-000-300-112 from Ministry of Education of Republic of Singapore.  J.-S. W. acknowledges support from a URC research grant R-144-000-257-112 of NUS. J.-T. L. acknowledges the Lundbeck Foundation for financial support (R49-A5454).

\end{document}